# CONTACT TRACING

*An Overview of Technologies and Cyber Risks*

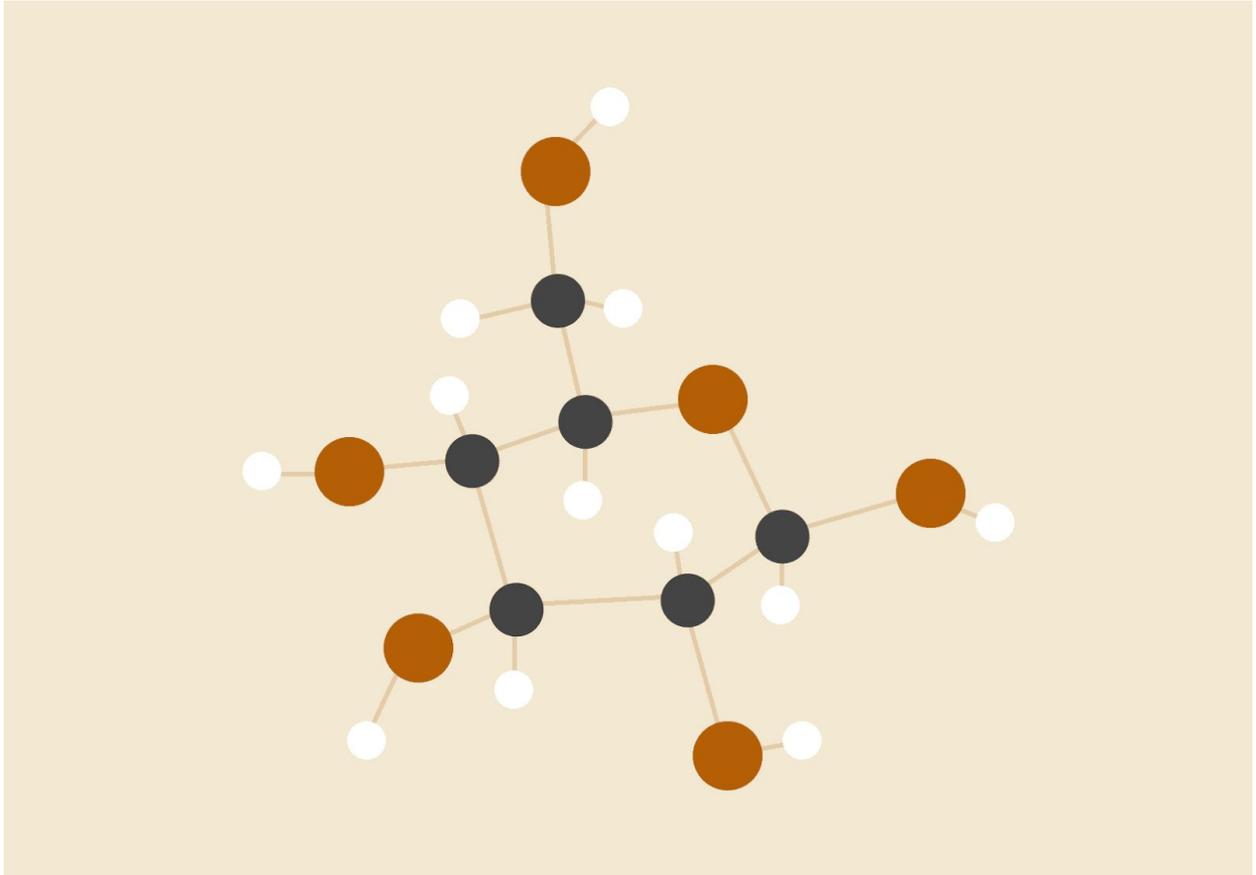

**Franck Legendre, Mathias Humbert, Alain Mermoud, Vincent Lenders**

armasuisse, Science and Technology, Switzerland

Corresponding author: vincent.lenders@armasuisse.ch

# EXECUTIVE SUMMARY

*The 2020 COVID-19 pandemic has led to a global lockdown with severe health and economical consequences. As a result, authorities around the globe have expressed their needs for better tools to monitor the spread of the virus and to support human labor. Researchers and technology companies such as Google and Apple have offered to develop such tools in the form of contact tracing applications running on smartphones. The goal of these applications is to continuously track people's proximity and to make the smartphone users aware if they have ever been in contact with positively diagnosed people, so that they could self-quarantine and possibly have an infection test.*

*A fundamental challenge with these smartphone-based contact tracing technologies is to ensure the security and privacy of their users. Moving from manual to smartphone-based contact tracing creates new cyber risks that could suddenly affect the entire population. Major risks include for example the abuse of the people's private data by companies and/or authorities, or the spreading of wrong alerts by malicious users in order to force individuals to go into quarantine. The Swiss population is concerned about these new risks and, according to a recent ZHAW study, 40% of interviewees fear that a smartphone-based contact tracing system could be turned into mass surveillance and believe that such an approach will simply not work with too many false notifications.*

*In April 2020, the Pan-European Privacy-Preserving Proximity Tracing (PEPP-PT) was announced with the goal to develop and evaluate secure solutions for European countries. However, after a while, several team members left this consortium and created DP-3T which has led to an international debate among the experts on how to securely implement contact tracing applications. At this time, it is confusing for the non-expert to follow this debate; this report aims to shed light on the various proposed technologies by providing an objective assessment of the cybersecurity and privacy risks. We first review the state-of-the-art in digital contact tracing technologies and then explore the risk-utility trade-offs of the techniques proposed for COVID-19. We focus specifically on the technologies that are already adopted by certain countries.*

# INTRODUCTION

This report describes different contact tracing technologies currently deployed and how they can help in containing the spread of COVID-19. The target audience is the general public that is concerned with the potential cyber risks posed by the use of those technologies.

Along with physical distancing and washing hands regularly, technology can help automate the manual work of contact tracing and slow down the spread of the virus. It can bridge the gap until the entire population reaches the so-called herd immunity or vaccines and treatments are available. End of April 2020, the infection rate among Geneva, CH inhabitants was 10% while it would require 65% to reach herd immunity. Experts say effective vaccine and antiviral treatments will be available only at the end 2020 / beginning of 2021.

With 92% of all adults in Switzerland owning a smartphone [Deloitte, Statista], digital contact tracing is seen as a complement to other measures to release the pressure on health services while restoring freedom of movement and restarting the economy.

## How Manual Contact Tracing Works

Manual contact tracing allows tracing back a potential chain of infections and giving early warnings to potentially infected people. This starts with Bob being diagnosed with COVID-19. He is then interviewed by a health inspector asking all his whereabouts and encounters from recollection of the last 14 days. The inspector then assesses all the potential contacts at risk who are then contacted to stay in quarantine and potentially take a test.

## How Can Technology Help Automate Contact Tracing

Manual contact tracing is a fastidious and slow process. This results in the virus always being one step ahead of the health inspectors. In order to contain the spread of the virus, a fast reaction time is key. Technology can help by providing immediate notifications to potentially infected persons by automating the contact tracing process, especially between commuters using public transports, or neighbours at a restaurant who would have no way to contact each other otherwise.

A study by Oxford epidemiologists published in Science concludes that with 80% application adoption (~57% of the entire population in the UK) and a clear test and quarantine policy, the virus can be fully contained. The ideal course of action is to notify contacts of a potentially infected person to self-quarantine as soon as that person has symptoms confirmed by health authorities. Those same contacts should be notified again as soon as the person has been tested positive or negative to recommend staying quarantined or be released, respectively. This strategy can contain the spread of the virus while minimizing quarantine time and scope given a high enough adoption.

## What Are The Risks of Digital Contact Tracing

Regardless of the technology being used for automating contact tracing, they all create new risks:

- **Privacy Risks** - from disclosing the identities of users infected by COVID-19 and their whereabouts, to revealing the real-world social network of an individual or part of the population.
- **Cybersecurity Risks** - mostly abusing the system to target specific individuals or companies with false notifications leading to unnecessary quarantine.

In this report, we review the different technologies which can be used for automating contact tracing. We assess their pros and cons and focus on the privacy and security risks they pose to individuals, companies and states. We conclude this report with our assessment of the existing solutions put in place, and we provide some guidelines to further mitigate the risks. All references are directly embedded as links in this report (see underlined text).

## TECHNOLOGY OVERVIEW

Different technologies can be used for automating contact tracing. In order to work, contact tracing needs to know that two persons were close-by for a certain amount of time. To infer people were close-by, one can use the smartphone's absolute location or relative location (proximity) to other smartphones. With absolute location, two smartphones are in contact if their geographic coordinates are within a predefined distance (e.g. 5m). With proximity technologies such as Bluetooth and WiFi, two smartphones will consider to be in contact when they can hear the signal of the other. Below we present in more detail the different technologies:

1. **Mobile Operator Contact Tracing -** The location of a mobile phone can be determined on the mobile operator side using the mobile operator's infrastructure. Multilateration of radio signals between cell towers can locate a phone with an accuracy of +/-140m in urban areas and up to kilometers in rural areas. The main advantage of the technology is that it is non-intrusive and can be put in place without any user intervention assuming a legal framework is in place. When applied to contact tracing, the main drawback is the poor accuracy and the serious privacy concerns that entail mapping a diagnosed individual's location trail with all the other individuals' trail who have crossed paths.

2. **Location-based Contact Tracing -** Smartphones can locate themselves using their on-device capabilities. Those include GPS for precise location, which however mostly works outdoors (+/-2m). For indoors where most encounters happen, device-side cell tower multilateration and crowd-sourced WiFi localisation (+/-10m) can be used. With newer WiFi access-points, indoor WiFi multilateration brings the accuracy down to 1-2 meters. Those capabilities combined have the main advantage of being more accurate than multilateration performed by the mobile operators alone but have the main drawback of requiring users to install a dedicated application on their phone.

3. **Proximity-based Contact Tracing -** While location-based contact tracing requires an absolute geographical location, technologies such as Bluetooth and WiFi allow inferring the relative proximity of smartphones by transmitting a small-range signal that others can hear and record (up to 50m outdoors and 25m indoors for Bluetooth). Those technologies have the main advantage of not having to disclose one's absolute location and offer a finer estimation of distance, especially indoors. It shares the similar drawback of location-based contact tracing requiring users to install an application.

Other approaches such as mobile or credit-card payments can be used to trace back contacts based on linking purchases from different customers around the same time in the same venue. This approach, used in South Korea, is not covered here.

## PRIVACY AND CYBERSECURITY RISKS

Adoption is key for the success of tracing applications and large adoption will be possible only if there is public trust in the developed application. In order to build this trust, privacy and security will have to be guaranteed, or at least maximized. There exist several different architecture models, location-based or Bluetooth-based, centralised or decentralised, and none of them can guarantee zero re-identification risk and full privacy for every user, especially those that become infected. However, this being said, different approaches will have diverse effects on privacy and security, and our goal here is to clarify the potential risks that each approach will create towards privacy and security.

In general, basic data protection principles such as *data minimization* (sharing the minimal amount necessary for the service's purpose), *purpose* or *storage limitation* (limit the use of data to the service's clearly defined purpose and limit its storage in time to the period the service is provided) should be guaranteed by the contact tracing application (GDPR). This means that the system should not learn more information about individuals than what it needs to fulfill the functional requirements of contact tracing (CT).

### Threat Model

For each of the following risks, it is also important to specify for each approach who is the potential *attacker* (who could jeopardize privacy) and who is the potential *victim* (whose privacy might be at risk).

In terms of attackers, there are essentially two main actors that we consider in our analysis. First, the central server, that can have access to more or less information depending on the CT approach. This central server may be run by the health authority. Second, any malicious third party, without any specific privilege, could try to gain information, e.g., by eavesdropping the CT-related communications, or to generate false alarms. Other potential attackers include the operating system, the network provider, the app developer, or the cloud provider, if any.

Among potential victims, we can identify three main categories: (i) the infected users, more precisely those who have been diagnosed positive and report that they have been diagnosed positive, (ii) the users who have been in contact with one or more infected users (exposed users), and (iii) any other users of the CT system.

### Privacy Risks

The risks on individual privacy can be categorised as follows:

1. **Health Status Privacy** - In general, independently of the type of contact tracing system, the first risk is to leak the identities of **users infected by COVID-19**. This information is by definition highly sensitive and is protected by medical secrecy. Therefore, it should remain accessible only to the infected users and the health authority. A related privacy risk is to learn the identities of users who have been in contact with an infected user (exposed users).

2. **Location Privacy** - Another privacy risk is to learn a user's mobility traces. Locations visited by a user can reveal a lot about her, from her political and religious views to her social relationships [NYT, NSA, walk2friends]. No system should *a priori* need access to location data to perform contact tracing. Geolocation-based contact tracing systems do however require location to infer proximity. Moreover, Bluetooth-based approaches could also indirectly reveal location data due to co-location information [Olteanu2017] and local Bluetooth sniffing stations [corona-sniffer].

3. **Social Graph Privacy** - Learning a user's social graph represents the third main privacy concern. This can be learned either directly through proximity data between users (for Bluetooth-based systems) or by relying on location data (for location-based systems) [NSA, walk2friends]. The system does not need to know the global social graph to perform contact tracing, but only the contacts between infected users and other users (proximity/local graph). Knowing the social graphs of a significant number of users can be further used to de-anonymize these users. For instance, if an attacker has access to side channel information, such as online social networks, he can match it to the global social graph he has reconstructed with contact tracing and then re-identify the users in this graph [deanon].

## Cybersecurity Risks

The cybersecurity risks are:

1. **False Alarms -** A user can create false alarms (e.g., through active relays and replay attacks). This could for example lead to the quarantine of employees of the defense, the government, or critical infrastructure.

2. **Passive Disruption -** A user can prevent notifications that other users are exposed, e.g., by (temporarily) disabling the Bluetooth or GPS functionality. Or simply by not reporting after diagnosis. This would reduce the traceability of the coronavirus and the ability to isolate infected users. This risk is inherent to any voluntary-based contact tracing system.

3. **Active Disruption -** An attacker can prevent contact discovery (e.g., by jamming the Bluetooth channel). This would reduce the traceability of the coronavirus and the ability to isolate infected users. This risk is inherent to any Bluetooth-based contact tracing system.

## INTERNATIONAL ADOPTION

When it comes to technology solutions, most countries have decided to use a proximity-based solution using Bluetooth Low Energy (BLE). We review below what solutions were chosen by early adopters and the learnings thus far. We also indicate what solutions have been chosen by other countries around Switzerland.

### Early Adopters

The table below shows the early adopters to digital contact tracing with the underlying technology used, the adoption and the main learnings.

|  | 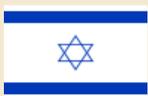 Israel | 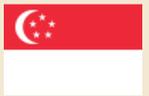 Singapore | 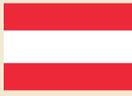 Austria | 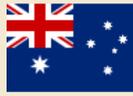 Australia |
|---|---|---|---|---|
| **Launch Date(s)** | 18/03/2020 | 20/03/2020 & 28/06/2020 | 01/04/2020 & 26/06/2020 | 26/04/2020 |
| **Technology** | Cell-phone location data processed by Shin Bet | TraceTogether App based on legacy BLE (OpenTrace) and TraceTogether Dongle for the elderly | Stopp Corona App based on legacy BLE initially and replaced by Apple-Google Exposure Notification (EN) (code repo) | COVIDSafe App based on legacy BLE (OpenTrace) |
| **Adoption (% of population[1])** | > 75% (every cell phone owner) | ~38% as of 02/07/2020 | ~7% as of 02/07/2020 | ~26% as of 23/05/2020 |
| **Main Learnings** | One third of Covid cases detected by the system (4,089 as of May 10th) with 80,072 alert notifications sent by SMS. | Background limitations of legacy BLE impose severe limitations on users as they need to keep their iPhones awake. | Low ratings on the Apple and Google App Stores due to too many invasive permissions of the first version (full analysis) led to the switch to EN. | 1. Suffers from the same limitations as TraceTogether 2. Australia is still considering switching to Apple-Google's Exposure Notification API. |

*Last updated: 02/07/2020.*

---

[1] Here we report the number of downloads, not the effective app usage among the population.

### Switzerland's Neighbors

Below we provide more details about the technology decisions of Switzerland and its neighbors with Austria already reported above:

|  | Italy | France | Germany | Switzerland |
|---|---|---|---|---|
| **Launch Date(s)** | 01/06/2020 | 02/06/2020 | 16/06/2020 | 25/06/2020 |
| **Technology** | Immuni App based on Apple-Google Exposure Notification (code repo) | StopCOVID ROBERT (centralised based on legacy BLE) (code repo) | Corona-Warn App based on Apple-Google Exposure Notification (code repo) | SwissCovid DP-3T and Apple-Google Exposure Notification (code repo) |
| **Adoption (% of population[2])** | ~7% as of 02/07/2020 | ~3% as of 23/06/2020 | ~16% as of 24/06/2020 | ~11% as of 01/07/2020 |
| **Main Observations and Learnings** | 3 people tested positive could alert other users on June 15th. | 1. App approved by the French Parliament and Senate 2. Only 14 alerts sent out | Initial design based on centralised PEPP-PT approach was dismissed | 1. App approved by the Parliament 2. 1 month trial phase |

*Last updated: 02/07/2020.*

NHSx in the UK decided to investigate both Bluetooth approaches (legacy and Apple-Google's EN) and after poor results with the legacy approach during trials on the Isle of Wight, the government announced on June 18th they will release an app using the Apple-Google approach (more details to the story) but not before winter.

Given the technical limitations of legacy Bluetooth, Apple and Google have partnered to provide a dedicated solution known as Apple-Google Exposure Notification API, which is more battery efficient and has no background limitations. More and more countries are adopting this *de facto* standard (Spain, Austria). This will provide a compatible solution between a few of Switzerland's neighbors (Italy, Germany, Austria) and an incompatible solution with France.

For a more detailed evaluation of the worldwide adoption, see this article from the MIT Technology Review.

---

[2] Here we report the number of downloads, not the effective app usage among the population.

## MOBILE OPERATOR CONTACT TRACING

Cell-phone location data is being used in Israel by the Shin Bet in collaboration with the Health department. It is the first government using such data for public health purposes. It required passing an emergency law on March 16th, 2020 to track people infected with COVID-19 including to identify and quarantine others they have come into contact with. *"Once an individual is highlighted as a possible coronavirus case, the health ministry will then be able to track whether or not they are adhering to quarantine rules. It can also send a text message to people who may have come into contact with them before symptoms emerged"*, according to the BBC. All Israelis possessing a cell phone are tracked through this system.

In Switzerland, such location data from Swisscom's Mobility Insights is used by the FOPH to *"see what impact the federal government measures have had on people heading to parks, popular tourist spots and other public spaces."* Such location data at a resolution of 100m x 100m -- aggregated and anonymized -- are already used to provide insights for urban or transport planning.

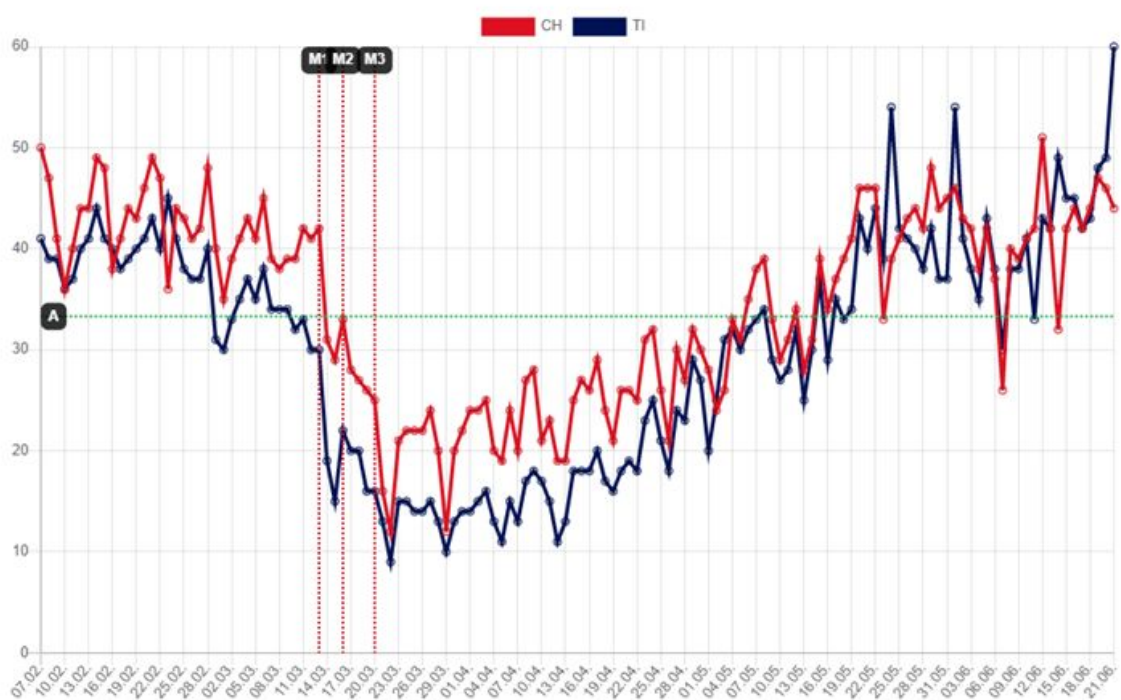

*Figure 1* - *Average number of kilometers travelled by Swiss overall (red) and specifically Ticinesi (blue) before and after the different federal lockdown measures (M1,M2,M3). Courtesy of Alain Jörg, Swisscom Mobility Insights.*

The low accuracy of mobile location tracking (+/-140m) makes it unfit for tracing contacts between specific individuals. That said, for the post-lockdown, this technology could be used in two ways:

1. **Impact of lockdown measures -** Assess the average density of people over time at public places like train stations or parks to assess lockdown measures are well respected.
2. **Environmental contamination -** Pinpoint potentially contaminated venues by looking for common places visited by infected individuals.[3] This would however require deanonymizing users' identity, which is not permitted by law in CH. If permitted, it could allow contacting the visitors of such venues. As an alternative, it could signal that a particular venue requires to be disinfected.

**Privacy/cybersecurity risks -** The approach of Israel is certainly the worst in terms of privacy risks. Indeed, the central server (be it the Shin Bet or the health ministry) can track all movements of all mobile users. This means it can then also know the infection status of mobile users, track their locations, and reconstruct their social graphs. Moreover, it is *not* on a voluntary basis, but mandatory by law. On the contrary, Swisscom's Mobility Insights provides only aggregated statistics to the FOPH (Federal Office of Public Health), which significantly reduces the risks of individual location tracking. The location data resolution was of 100m x 100m and the released statistics contained at least 20 individual records in order to prevent re-identification attacks. Cybersecurity risks are clearly minimized with the mobile operator contact tracing approach since it controls the entire system.

**Main Pros**

- Non-intrusive - users do not need to opt-in or install any application.
- High adoption - with a single mobile operator in most countries.
- Government oversight of notifications to exposed individuals.
- Cybersecurity risks are minimal.
- Privacy risks of Mobility Insights are limited thanks to statistics aggregation.

**Main Cons**

- Limited accuracy makes it unfit for contact tracing between individuals.
- Critical mass - Swisscom with its market share covers 60% of the population and has a critical mass for data to be exploitable. In other countries with no dominant operator, cooperation between operators would be required.
- The Shin Bet approach is the most privacy invasive among all contact tracing technologies in modern democratic countries.

---

[3] According to Ferreti et al, environmental contamination accounts for less than 10% of infections by COVID-19.

## LOCATION-BASED CONTACT TRACING

A few countries such as Iceland and India decided to use location-based applications. We review here SafePath from MIT, which strives to offer a privacy-preserving solution for such applications. We are not aware of any country using this approach already.

SafePath uses the smartphone location capabilities in a way that attempts to preserve privacy, as described in Figure 2 below. It enables individuals to log their own location called "trails" on their device. Once diagnosed positive, individuals can provide health officials with accurate location trails (1). The health authorities receiving those trails are equipped with a tool to remove sensitive location trails for diagnosed carriers and local businesses (2). This sanitized location trail is then broadcast to all other SafePath users (3), who can compare it with their own trails and determine whether they have crossed paths with the diagnosed carrier (4).

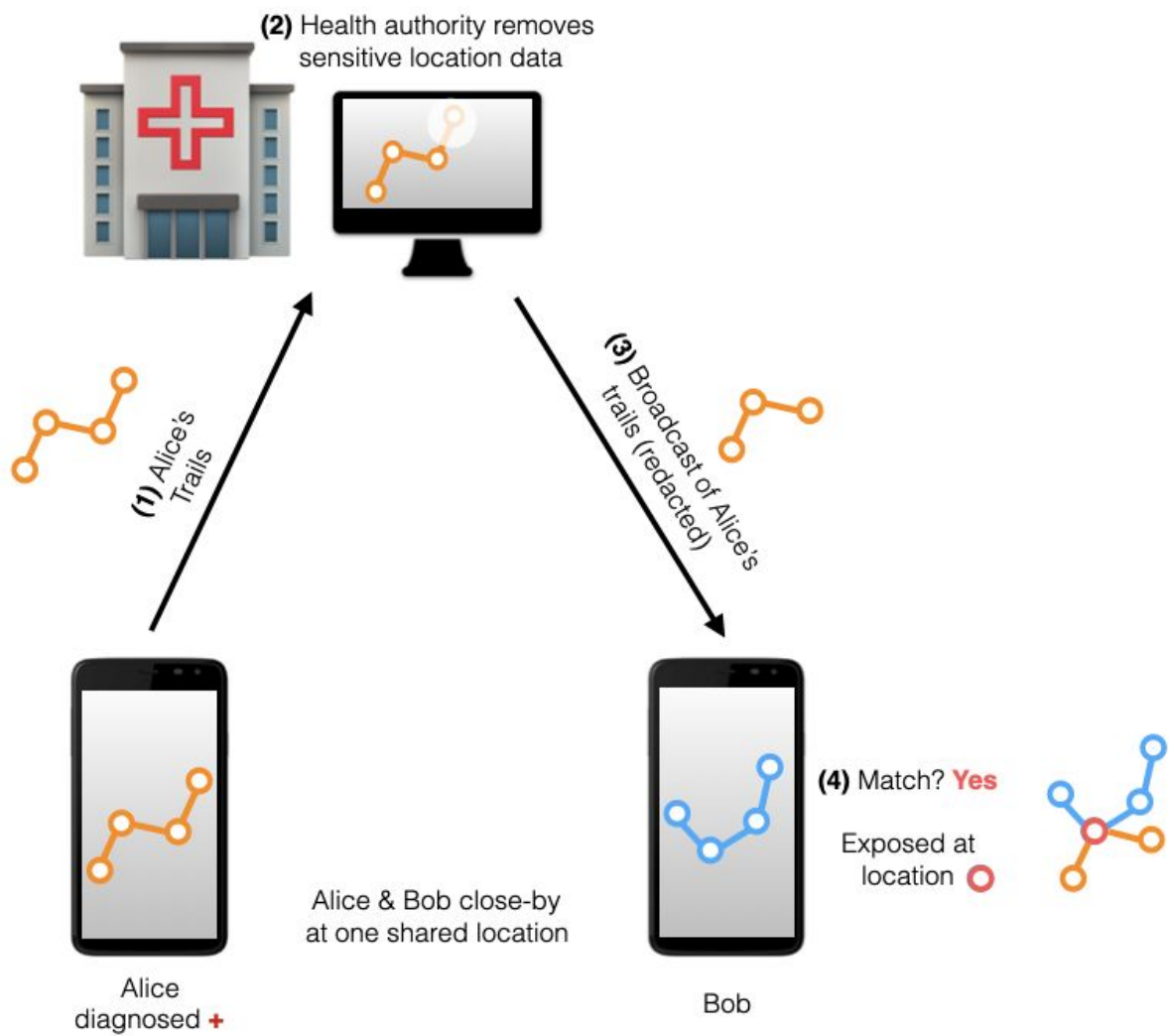

*Figure 2 - SafePath (first version).*

**Privacy/cybersecurity risks -** SafePath's approach is one step ahead in terms of privacy compared to other GPS-based solutions or the one developed by the Shin Bet. Safe Path's ability to do the matching on the device, that is, without collecting the trails from all users, prevents massive government surveillance. However, its first two versions still allow the central server/health authority to have access to the location trails of all infected users, including sensitive location data. This will also reveal co-location between infected users, and thus part of the social graph. The central server can also learn who has been infected. An upcoming version of SafePath should also encrypt the sanitized location traces before sending and comparing them (encrypted). This is a promising direction but it still lacks details to be fully evaluated. There are also high cybersecurity risks due to GPS spoofing attacks (which can simply be performed with an application, or with more advanced techniques to take over the GPS signal seamlessly [Tippenhauer11]). In such an attack, a malicious user can send a fake GPS signal and make other users think she was close to them. This could be used to generate numerous false alarms, e.g., towards targeted employees or governmental officials.

**Main Pros**

- Privacy-first approach with decentralised processing of trail matching.
- Enables to assess environmental contamination.

**Main Cons**

- Low precision, especially indoors, which might lead to both high false positives and false negatives.
- The ability to fake the GPS location of the users opens the door to serious cybersecurity risks (false alarms).
- Even though privacy risks are reduced with SafePath compared to other mobile operator or location-based approaches, in its first versions location traces are leaked to the central server, which raises serious privacy concerns.
- Users have to trust the health authority to implement the privacy mechanism.

## PROXIMITY-BASED CONTACT TRACING

In both centralised (ROBERT, FR) and decentralised (DP-3T and Apple-Google Exposure Notification, CH) approaches, users regularly exchange ephemeral identifiers with other users nearby. In order to exchange their identifiers, both users need to have the contact tracing application installed and their Bluetooth activated. Those identifiers are depicted with fruits in Figure 3 below at steps (1) and (2). The ephemeral identifiers are regularly updated (e.g., every 15 minutes). The application sends to and receives identifiers from other nearby users. It also stores two lists: (i) one for the sent identifiers, and (ii) one for the received identifiers (see tables in the figures below). These lists are regularly updated by removing old identifiers that are not useful for tracing the coronavirus infection anymore (typically those sent/received more than two weeks ago). The centralised and decentralised approaches mainly differ in how they generate the identifiers and how they identify exposed users.

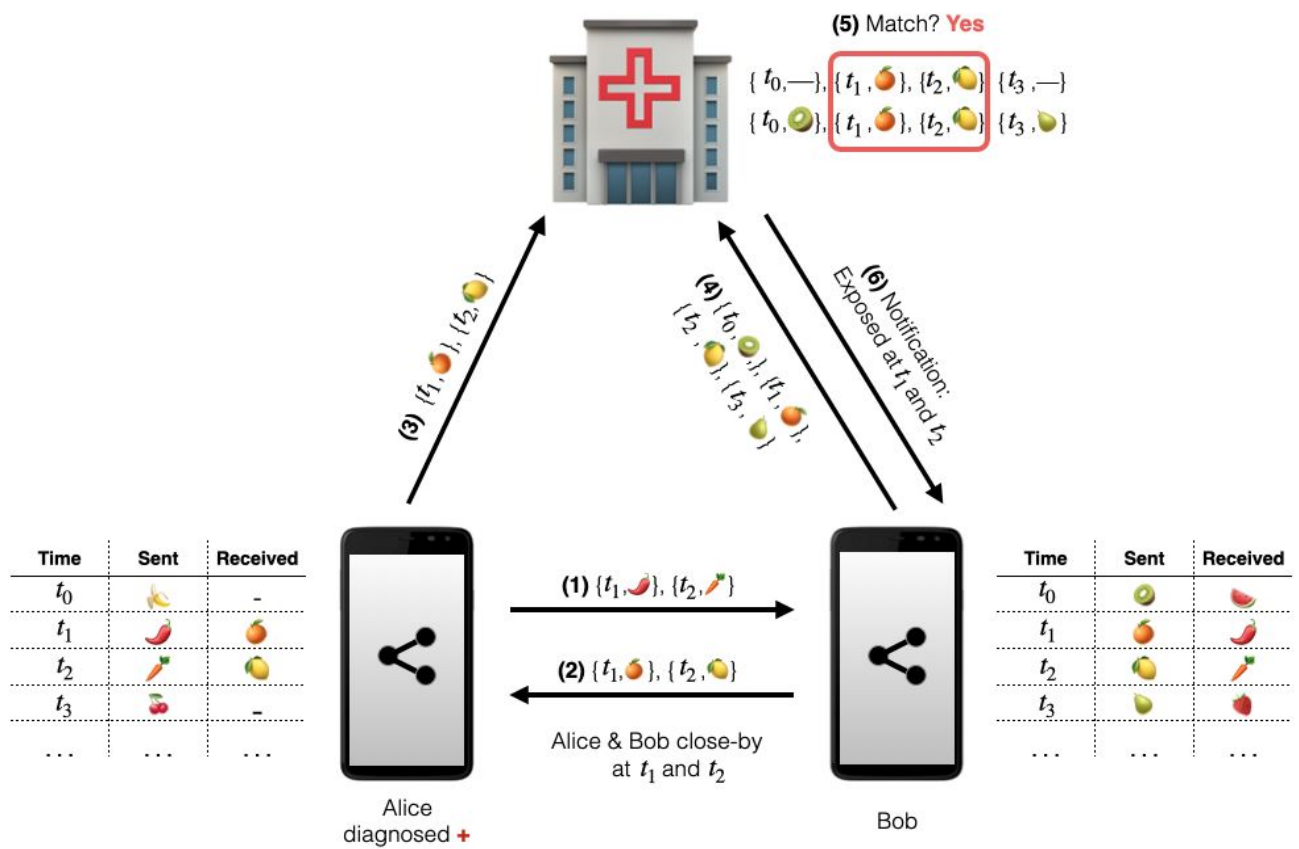

*Figure 3* - *Centralised Proximity-based Contact Tracing such as ROBERT.*

In centralised systems -- such as ROBERT -- the ephemeral identifiers are generated by the server and then sent to the app. Moreover, they are linked to a long-term pseudonym. Upon positive diagnosis, in centralised systems the infected user sends the list of *received* identifiers to the server (3). The app of other users queries the server to know whether the user has been in contact with an infected user (4 & 5). Indeed, the server stores the list of users the infected user has encountered and can map it to a long-term pseudonym. Therefore, the server can easily check whether any user has been exposed and notify the app accordingly (6).

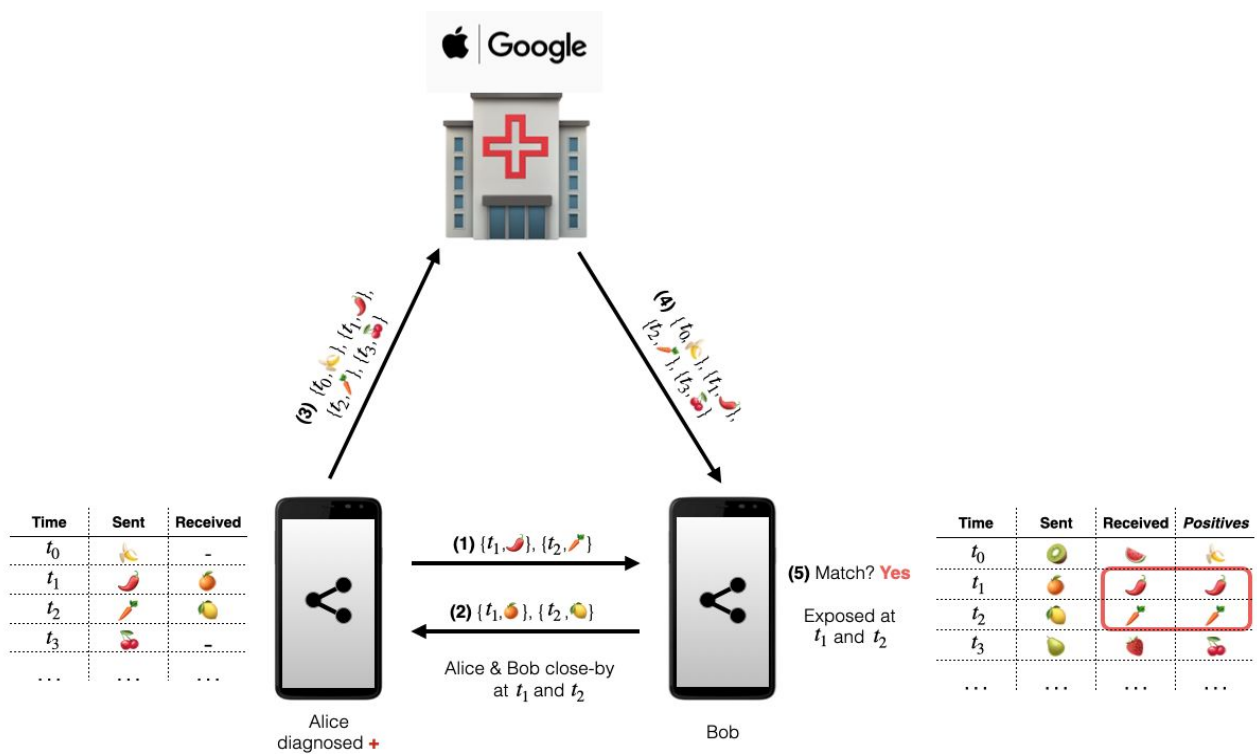

*Figure 4* - *Decentralised Proximity-based Contact Tracing such as DP-3T and Apple-Google's Exposure Notification.*

In decentralised systems -- such as DP-3T and Apple-Google Exposure Notification -- the ephemeral identifiers are directly generated by the app of the user (or the OS). When diagnosed positive, the app sends the list of *sent* identifiers to the server (3). In practice, a condensed form of those identifiers known as the diagnosis key is sent to the server. In order to determine whether a user has been in contact with infected users (i.e., exposed), Bob app downloads the list (of sent identifiers) of infected users (4) and checks *locally* if one of the received identifiers (stored by the app) is part of the downloaded list (5).

**Centralised vs. Decentralised** - Besides the differences on privacy, the centralised approaches have the following main advantages compared to the decentralised approaches:

- **Pandemic Oversight** - By centralising data, the centralised approach gives the health authorities a better overview of the current spread of the virus in real-time (i.e. number of persons potentially at risk). It eases identifying clusters if many people report having seen the same identifiers ((4) in Fig. 4). It also provides more data to epidemiologists to model epidemics and fine-tune the computation of the risk factor (indicating the risk of each individual). This risk factor will be determined by parameters that can be fine-tuned based on data analysis, and help minimize false positives (i.e. help receiving too many false alarms/notifications) and false negatives (i.e. not receiving a notification when one should have). Note that the above is also possible with the decentralised approaches, but it is assumed that users will have to give their consent by opt-in to share their data with health authorities and epidemiologists.

- **Faster Notifications** - In the centralised approaches, the health authorities will be in charge of the servers and have therefore more control on notifications sent to persons deemed at risk. With the Apple-Google decentralised solution, diagnosis keys will be broadcast to all users every 24h only. This in order to gather enough diagnosis keys and prevent the re-identification of diagnosed individuals. This also means that the on-device matching to know whether one has been exposed will be delayed consequently. According to the Oxford study [Fraser], each single hour counts and delays in notifying contacts could jeopardize the containment of the virus.

**Privacy/cybersecurity risks -** We summarize here the findings of the detailed risk analysis carried out in the next section. Centralised systems tend to put at risk the privacy of *all users,* especially against the central server, while decentralised systems tend to put at risk the privacy of *infected people* against anyone (e.g., another user). Centralised systems put more location privacy and social graph privacy at risk than decentralised systems. However, decentralised systems tend to be more vulnerable to the re-identification of infected users. Besides that, the centralised approach provides a single point of failure that a successful hacker could leverage to gain access to all the information stored by the central server. It further gives the ability for authoritarian regimes (or any state with subpoenas) to look for a specific person (under investigation) and get access to all her data.

In terms of cybersecurity risks, proximity-based contract tracing is also prone to false alarm attacks, although these are more challenging to carry out than with location-based contact tracing. The decentralised approach would be particularly prone to this attack as no control is envisioned as opposed to a centralised approach where the health authority has oversight of notifications and can detect "mass notifications". To counteract such attacks, one would need to send blacklists of ephemeral IDs that should not be accounted for the calculation of the risk factor.

**Main Pros**

- Bluetooth can estimate immediate proximity more accurately than GPS-based approaches, especially indoor (see this study).

- Bluetooth works indoors and even offline.
- Privacy is better preserved relatively to other (mobile operator or location-based) approaches, especially in the decentralised scheme.

**Main Cons**

- Contact data gives no context about the environment (indoor vs. outdoor, public vs. private places), which epidemiologists could use to better understand how the virus spreads.
- The propagation of electromagnetic waves correlates poorly with that of droplets (for instance, it can cross walls), and it is hard to map noisy signal strength measurements to distances beyond 2m.
- Compared to mobile phone operator tracing, it needs the mobile phone user to install an app. Compared to location-based contact tracing, it needs the continuous activation of Bluetooth which exposes the users to the risks of being tracked by third parties with a Bluetooth sniffer.
- Although more challenging, cybersecurity attacks such as false alarms are still feasible.

## DETAILED RISK ANALYSIS OF PROXIMITY-BASED CONTACT TRACING

We review below in more details the privacy and cybersecurity risks of the two main Bluetooth-based contact tracing solutions, namely EN/DP-3T (decentralised) and ROBERT (centralised) since those are the two main systems being used worldwide. We do not cover the risks of recent proposals [Epione, Desire] since those solutions are mid-term to long-term solutions that cannot be readily available to the public.

### Risks on Health Status Privacy

**EN/DP-3T** - An attacker can identify the infected users he has been physically near to. He can do so by creating multiple accounts in the contact tracing systems and by using them for a short period of time. He can also do so by frequently rotating his own Bluetooth broadcast identifier to be able to later retrieve which infected user he has encountered, provided that he has not been in contact with multiple infected users during this timeframe. The attacker can match the set of infected identifiers against each of her recorded Bluetooth identifiers to determine when he was in contact with an infected person and use this information to reveal the identity of the infected user. Even though the ephemeral identifiers are set to be updated, e.g., every 15 minutes, this could be modified by the attacker to be updated more frequently [DP-3T analysis].

**ROBERT** - It is more difficult for an attacker (other than the central server) to identify infected users since the central server can make the registration of multiple dummy accounts more difficult. For instance, ROBERT currently requires a CAPTCHA to register an account, even though it is probably not sufficient to avoid sybil attacks at a large scale. However, with a centralised approach, the central server learns who is infected, but this will also certainly happen if the server is also the health authority. Moreover, the server can learn the identities of users observed by the infected user (i.e., exposed users). This is due to the fact that the server can associate ephemeral Bluetooth IDs to long-term identifiers. Besides the server itself, there is always the risk of unauthorized access to this data by third parties (e.g., hackers).

Note that in general there is always the risk, in both approaches, that the central server learns which users are infected by looking at their network identifiers (e.g., IP address). This could be mitigated by relying on a proxy, e.g., a local hospital which must be itself trusted, a VPN, or an anonymity network like Tor.

### Risks on Location Privacy

Despite the fact that proximity-based approaches do not collect location data, they are also prone to location privacy breaches under some conditions, as described below.

**EN/DP-3T**- there is a risk of location tracking by a malicious third party installing enough Bluetooth sniffers at different physical locations and a mobile device hacked (i.e. rooted or jailbroken) [corona-sniffer]. Here only infected users would be at risk of being tracked. It is very difficult to prevent such an attack against a global eavesdropper. See below for more details about the attack.

**ROBERT** - there is a risk of location tracking (by the central server) with the centralised approach if the central server has access to an auxiliary channel such as CCTV footage or local eavesdropping stations (sniffers) that enables it to associate a location to a pseudonymous long-term identifier. This attack would affect any user, i.e., both infected and non-infected users.

Example: Location Disclosure of Infected Users with Exposure Notification

The location of diagnosed individuals using the DP-3T or Apple-Google EN solution can be disclosed by collecting their transmitted Bluetooth signal. This attack detailed by Otto Seiskari works as follows. Bluetooth sniffers, potentially smartphones with a dedicated app, listen to the DP-3T or Apple-Google Exposure Notification transmitted by individuals. Those smartphones also record the location where those IDs were heard. This data can then be sent to a central database. Whenever someone is diagnosed positive and her key is broadcast to all users of the contact tracing application, a hacker can intercept this diagnosis key. From this key, one can derive the ephemeral IDs that were generated and query the database for the location where those IDs were seen. As shown by this simulation below, given enough sniffers, one could uncover the whereabouts of diagnosed individuals.

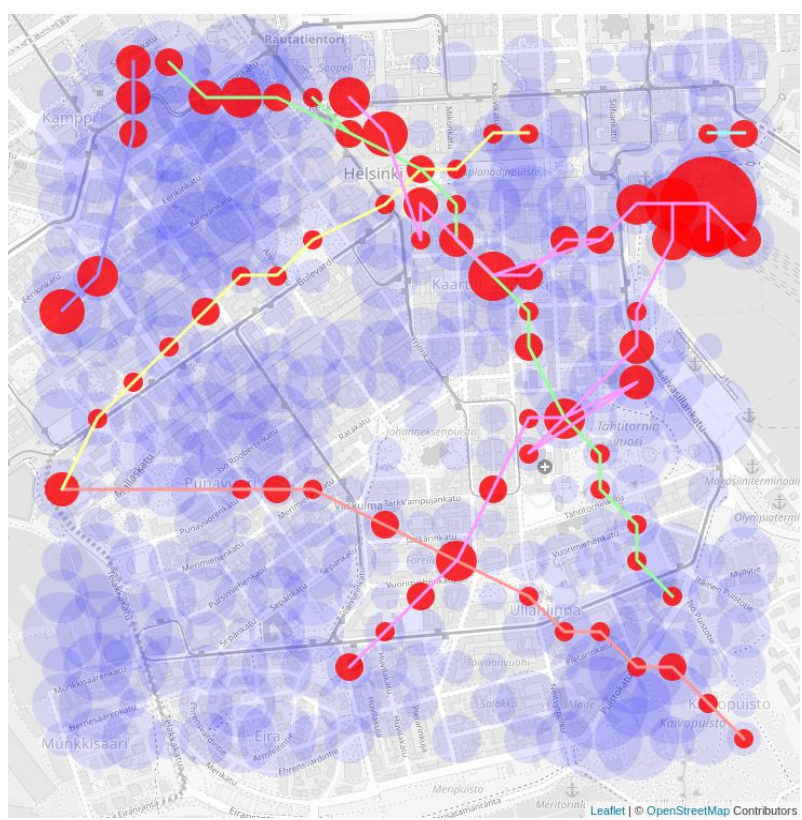

*Figure 5* - *Courtesy of Otto Seiskari and © OpenStreetMap*

Figure 5 shows the simulation results where 400 BLE-sniffing devices have been deployed over an area of 1500×1500m² with 300 individuals. The red circles correspond to the Bluetooth signals recorded from infected individuals, who have voluntarily uploaded their positive infection status to the local health authorities. Blue circles are signals recorded from other people using the contact tracing service. As illustrated by the lines connecting the red dots, the route traveled by each infected individual can be reconstructed within the range of the sniffer grid.

Even though the transmission of the diagnosis key will be secured and stored securely on the hardware keystore of iPhones and Android (Secure Enclave and Trust Platform Module on iPhones and Android, respectively), most Android phones sold before 2018 do not have this secure keystore. It only requires rooting one "unsecure" Android phone and extracting the key from memory to potentially uncover the locations visited by diagnosed individuals through this attack.

### Risks on Social Graph Privacy

**EN/DP-3T**: In the decentralised approach, the central server will only learn colocation information between infected users, and nothing more. This will dramatically reduce the coverage of the social graph, and certainly prevent further attacks based on it. This should represent limited information as long as only a very limited percentage of users get infected. Finally, a recent attack also showed that someone could formally prove that he encountered a person who was diagnosed as infected. This attack also affects health status privacy by making infected users easier to identify.

**ROBERT:** This is probably the most serious privacy risk with the centralised Bluetooth-based approach. The central server learns all the interactions of infected users which can quickly percolate to a significant portion of the social graph of all users (due to the structure of social graphs and small-world phenomenon). It also raises interdependent privacy risks, since it can indirectly expose the social graph of other, non-infected, users. It is well-known that once we have a significant portion of a social graph, it is possible to de-anonymize the graph with auxiliary information such as online social networks [deanon].

### Summary of Privacy Risks and Potential Improvements

In summary, centralised systems tend to put at risk the privacy of *all users*, especially against a malicious central server, while decentralised systems tend to put at risk the privacy of *infected people* against anyone, as also observed by [Vaudenay]. Moreover, centralised systems put more location privacy and social graph privacy at risk than decentralised systems. However, decentralised systems tend to be more vulnerable to re-identifying infected users. Besides that, the centralised approach provides a single point of failure that a successful hacker could leverage to gain access to all the information stored by the central server. It further gives the ability for authoritarian regimes (or any state with subpoenas) to look for a specific person (under investigation) and get access to all her data.

Note that privacy can be further enhanced using private set intersection, private information retrieval, or homomorphic encryption [Epione, Altuwaiyan18, Canetti20, Cho20], and prevent some of the above risks, such as learning the users' social graph or learning who has been diagnosed positive. For instance, by using two-party private set intersection-cardinality (PSI-CA), neither the central server nor other users can infer the infection status of the users [Epione].

### Cybersecurity Risks

Given that active and passive disruption is inherent to Bluetooth-based voluntary contact tracing and relatively simple to understand, we focus in the following on the risks of false alarms with a malicious intent (not because of the contact tracing technology).

The most threatening cybersecurity attack is the one that generates false alarms (either to targeted people or to large numbers of users) for various malicious purposes. One key threat is to send false alarms to employees of critical infrastructures (power plants, military bases, etc.). A typical scenario is one where the attacker can recruit someone with symptoms and get his phone. The attacker with the borrowed phone can then get in close contact with the targeted employees (e.g., in bars at night for soldiers). When the person with symptoms gets positively tested at the hospital, it will raise an alarm to all exposed employees and recommend them to stay in quarantine (until they get tested themselves). A more technically sophisticated approach is to relay Bluetooth signals of users that are or might be diagnosed soon (e.g., by staying close to a testing center). Fig. 6 below illustrates such an attack carried out by Dave and Bob. Bob stays close-by to Alice who is currently at a test center to get diagnosed. She will soon be diagnosed positive. Bob listens to the IDs that Alice is sending (1). He relays those over the Internet to Dave (2 and 3). Dave is close to his target and replays sending those IDs as if they were sent by Alice. When Alice is confirmed positive and the target receives Alice's diagnosis key from the health authority, it will receive a notification saying he has been exposed.

Such attacks can also be carried remotely, potentially even as far as from a satellite belonging to a rogue state that would relay those signals of infected or soon to be diagnosed individuals [4G over satellite, SMS from Satellite] (4 and 5). The decentralised approach would be particularly prone to this attack as opposed to a centralised approach where the health authority has oversight of notifications and can detect such attacks when done *en masse* (i.e. "mass notifications"). To counteract such attacks, one would need to send blacklists of ephemeral IDs that should not be accounted for the on-device calculation of the risk factor.

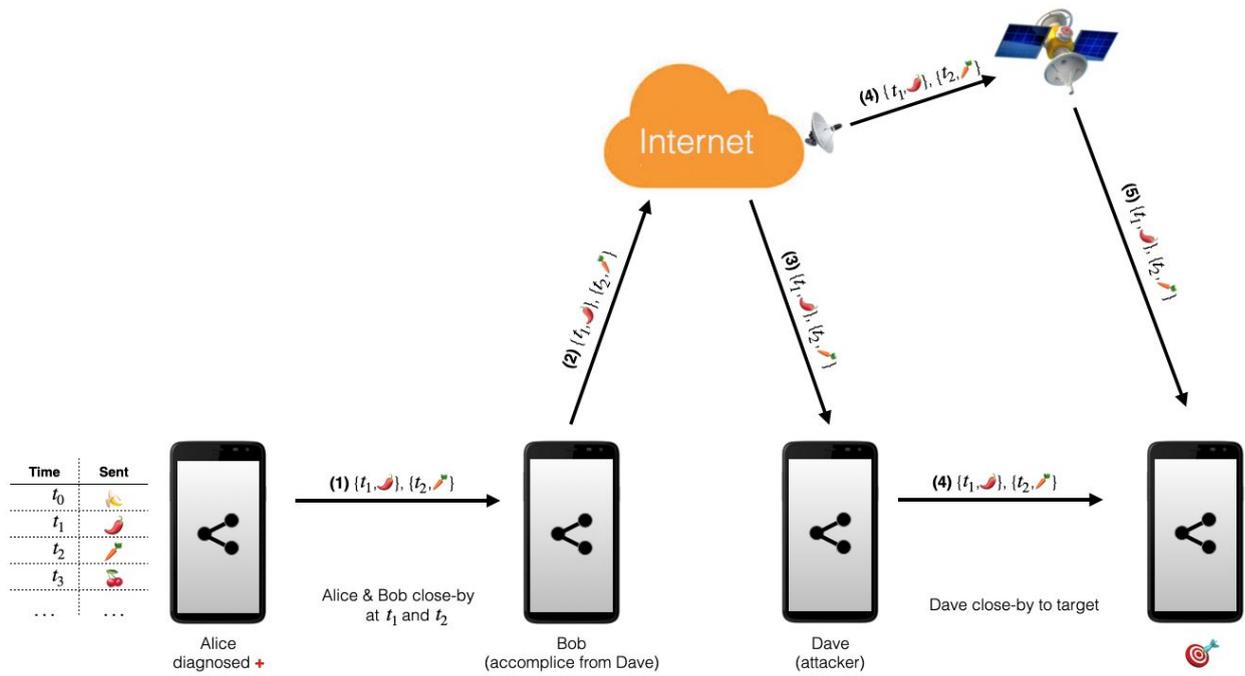

***Figure 6*** - *Examples of targeted attacks from (i) a close-by attacker with a phone (Dave), and (ii) from a remote attacker with a satellite, both relaying legit IDs from Alice diagnosed (or soon to be diagnosed) positive.*

As a final note, we would like to stress that the main difference between DP-3T and Apple-Google's Exposure Notification is that the Apple-Google solution is closed-source, and that Apple-Google may exploit the data for their own purpose.

## COMPARISON SUMMARY

In the table below, we summarize our assessment of the different solutions reviewed in this report using different criterias with a score from 1 (bad) to 5 (excellent). Note that these grades were derived by consensus among the authors of this review and are open to the debate. They could also change with newer versions of the technologies.

|  | Mobile phone Tracking | GPS Tracking App (SafePath) | Bluetooth Centralised Tracking App (ROBERT) | Bluetooth Decentralised Tracking App (Apple-Google, DP-3T) |
|---|---|---|---|---|
| Efficiency/accuracy (precision and notification time) | 3 (*) | 2 | 2 (**) | 3 |
| Privacy | 1 (*) | 2 | 3 | 4 |
| Cybersecurity | 4 | 2 | 4 | 3 |
| Battery efficiency | 5 | 3 | 3 | 4 (***) |
| Adoption likelihood | - | 1 | 2 | 3 |
| **OVERALL SCORE** | **3.3** | **2** | **2.8** | **3.4** |

(*) These numbers relate to the Israelian approach, not Swisscom's Mobility Insights.

(**) The centralised approach has a lower score despite its advantages mentioned earlier (i.e. pandemic oversight and faster notification, see Centralised vs. Decentralised in PROXIMITY-BASED CONTACT TRACING) because it can only rely on legacy BLE, which has shown to be very unreliable for peer discovery on iPhones (see INTERNATIONAL ADOPTION) compared to the dedicated Exposure Notification offered by Apple-Google.

(***) Apple-Google's Exposure Notification is expected to be more power-efficient than other solutions using the legacy Bluetooth API since it is handled at the OS level taking advantage of a phone's duty cycles.

## CONCLUSION

The hype around digital contact tracing started with scientific studies demonstrating that Bluetooth-based solutions could prevent a second wave if enough users participate. But, in practice, there are too many unknowns.

In terms of effectiveness, it is still not proven that Bluetooth can provide an accurate estimation of distance and not result in too many false alarms [Dehaye]. Also, given that most countries will rely on Apple-Google's solution, which does not provide the raw data to estimate such distance, it is questionable how the Apple-Google solution can be improved to reduce the false positives and false negatives as epidemiologists will have little data to work on.

The current version of the decentralised approaches adopted by many countries (incl. Switzerland) puts the privacy of infected individuals at risk. Centralized approaches such as those adopted by France with ROBERT tend to put at risk the privacy of *all users*, especially against a malicious central authority or a hacker targeting this authority. In terms of cybersecurity, all app-based solutions, whether location-based or proximity-based, are prone to targeted attacks. A location-based app can be tricked with GPS spoofing where an attacker can fake her location and target a location from anywhere on earth and pretend it was close to its victim. Although more difficult, Bluetooth-based approaches are also prone to such attacks where an attacker can relay "infected" IDs to a target by being close-by, but also remotely with satellite communication.

Given the privacy and cybersecurity risks and an effectiveness that is still to be proven, we conclude that there isn't any technological solution yet that can fulfill effectiveness while providing privacy guarantees and being cyber-attack free.

Shall a contact tracing solution be launched, we provide the following recommendations to mitigate targeted cyber-attacks and also how to react in case of attack or suspicion of attack:

1. Digital contact tracing is mainly of value to trace back strangers in public transport and spaces or, e.g., neighbours at a restaurant where we would have no way to contact otherwise. It is questionable why this should be an "always-on" technology. We recommend disabling Bluetooth in private spaces where manual contact tracing is in place.
2. Consider any smartphone left unattended as suspicious (or any other small battery-powered computation device such as a Raspberry Pi).
3. Many alarms received by a subset of related people concomitantly might be a sign of an attack.

It is important to also note that any legal recourse against attackers will be made difficult as the identity of an infected accomplices sharing its infecting Bluetooth beacons cannot technically be disclosed in decentralised approaches.

For the future and the next pandemics, many improvements can be made to render digital contact tracing effective. Regarding accuracy, 5G combined to other on-device location capabilities will provide a higher precision (although only in urban areas). For privacy, many solutions have already been proposed to provide higher guarantees using for example secure multi-party computation (such as private set intersection). But those, while providing stronger privacy guarantees, do not address all cybersecurity risks.

## ACKNOWLEDGMENTS

A special thank you goes to the external reviewers for their insightful feedback which helped us enhance this report:

Panayotis Antoniadis, Co-founder NetHood Zurich
Domenico Giustiniano, Research Associate Professor at IMDEA Networks
Julien Herzen, Engagement Director, Unit8 SA
Kévin Huguenin, Full Professor, UNIL-HEC Lausanne
Dimitri Percia David, Cyber-Defence Campus Distinguished Postdoctoral Fellow, University of Geneva
Damian Pfammatter, Scientific Project Manager, Cyber-Defence Campus, armasuisse S+T
Eiko Yoneki, Affiliated Lecturer and Senior Researcher leading Data Centric Systems Group in the University of Cambridge Computer Laboratory Systems Research Group